# Flood Evacuation During Pandemic: A multi-objective Framework to Handle Compound Hazard

Shrabani S. Tripathy[1], Bhatia Udit[2], Mohit Mohanty[3], Subhankar Karmakar[1, 3, 4], Subimal Ghosh[1, 4, 5, *]

[1]Inter-Disciplinary Program in Climate Studies, Indian Institute of Technology Bombay, Mumbai – 400076, India

[2]Dept. of Civil Engineering, Indian Institute of Technology, Gandhinagar, Palaj – 382355, India

[3]Environemntal Science and Engineering Department, Indian Institute of Technology Bombay, Mumbai – 400076, India

[4]Centre for Urban Science and Engineering, Indian Institute of Technology Bombay, Mumbai – 400076, India

[5]Department of Civil Engineering, Indian Institute of Technology Bombay, Mumbai – 400076, India

[*]Correspondence to Subimal Ghosh, Email: subimal@civil.iitb.ac.in

## Abstract

The evacuation of the population from flood-affected regions is a non-structural measure to mitigate flood hazards. Shelters used for this purpose usually accommodate a large number of flood evacuees for a temporary period. Floods during pandemic result in a compound hazard. Evacuations under such situations are difficult to plan as social distancing is nearly impossible in the highly crowded shelters. This results in a multi-objective problem with conflicting objectives of maximizing the number of evacuees from flood-prone regions and minimizing the number of infections at the end of the shelter's stay. To the best of our knowledge, such a problem is yet to be explored in literature. Here we develop a simulation-optimization framework, where multiple objectives are handled with a max-min approach. The simulation model consists of an extended Susceptible - Exposed - Infectious - Recovered – Susceptible (SEIRS) model. We apply the proposed model to the flood-prone Jagatsinghpur district in the state of Odisha, India. We find that the proposed approach can provide an estimate of people required to be evacuated from individual flood-prone villages to reduce flood hazards during the pandemic. At the same time, this does not result in an uncontrolled number of new infections. The proposed approach can generalize to different regions and can provide a framework to stakeholders to manage conflicting objectives in disaster management planning and to handle compound hazards.

## 1. Introduction

COVID-19 is a highly infectious respiratory disease, first identified in December 2019 in Wuhan, China. It spreads through small droplets while talking, coughing, or sneezing and has been declared a global pandemic by the World Health Organisation (WHO) (https://www.who.int/emergencies/diseases/novel-coronavirus-2019/question-and-answers-hub/q-a-detail/q-a-coronaviruses; last access: 24 September 2020). Covering face, washing hands frequently, and social distancing are the essential preventive measures that should be taken in order to curtail the spread. It is challenging to continue such preventive measures during the occurrences of natural hazards, including floods and cyclones. The risk due to the pandemic along with the hydro-meteorological hazard together emerges into compound risk. Phillips et al. (1) have highlighted the possibility of such compound hazard occurrences during a pandemic. The likely occurrences of flood over Eastern India during the pandemic is considered as one of the potential examples.

Flood has been one of the most devastating natural disasters that cause massive loss of lives and property (2,3). Adequate preparedness and disaster management planning are required to minimize these losses and increase recovery speed (4). During floods, evacuation is one of the most critical preparedness measures to minimize the loss of lives, where people from high flood risk areas are shifted to safer areas (5). The objective of evacuation planning is to define a policy for people under high risk to minimize loss of lives and damage to property (6).

Preparedness for floods and cyclones starts by creating safe shelters at strategic locations, not very far from the high hazard areas. The evacuation planning involves shifting the vulnerable populations efficiently to the shelter homes while ensuring the timely distribution of essential commodities (7). Preparing an evacuation strategy well before the flood occurrence is pivotal to avoid last moment chaos that occurs due to the involvement of decision-makers at multiple stages and the need for the necessary arrangements to implement the evacuation in real-time. Informing the evacuees well in advance about the evacuation will ease the process largely and make the evacuee comfortable in following the instructions (8).

Optimal allocation of evacuees to shelters is a key challenge in evacuation planning (4). Under normal circumstances, the only objective is to decrease the number of people under flood risk, so as to maximize the number of people to be evacuated to nearby shelters. These shelter homes are designed to accommodate a very large number of people (for example, the capacity of a shelter on the East coast of India is approximately 2000) during natural disasters. As the shelters are provided for a very short duration (around one week), the per capita area allocated is low (9). This is acceptable during normal scenarios; however, during the pandemic, it is essential to maintain social distancing to control the spread of COVID-19. Hence, during the pandemic scenario, it is not desired to fill the shelters at their full capacity. On the other hand, the evacuation demands for the shifting of maximum people to the shelters from the (possible) flood-affected regions. The two objectives, to reduce the spread of pandemic (COVID-19 here) and increase the number of evacuated people, are in conflict with each other. This poses a challenge to disaster mitigation organizations and policymakers.

People under potential risk are evacuated to safer shelter houses timely and safely (5,10). Evacuation planning involves a number of decision-makers and disparate individual behavior of evacuees. An effective evacuation planning requires well-defined roles, responsibilities, and communication amongst stakeholders (11). Evacuation planning depends on factors like geographical location, population size, the spatial extent of the event's extremes, duration, the intensity of the event, and uncertainties (12–14). Understanding the evacuation process and the associated models are necessary for evacuation planning (15). Mathematical modeling and optimization have become helpful tools for evaluating time requirements for evacuation and allocating evacuees in optimal shelters (7,16). Various studies have used optimization models for flood evacuation to minimize losses considering factors like travel time and distance, cost of evacuation, and usage of infrastructure (5,6,14,17–19). Most of these studies have considered the objective function as the minimization of the transportation distance and/or time required to reach the shelters.

While the objective of designed evacuation strategies is to minimize the injuries and loss of life during the disaster, the prevalence of contagious diseases, including COVID-19, present conflicting priorities to the stakeholders and policymakers. Flood evacuation strategies are designed to encourage people to take shelters in designated areas. However, violation of social distancing protocols in these shelters could result in a sudden surge in the contractions of infections and mortality rates (20). Besides, immediately following a disaster and throughout the recovery period, healthcare facilities are often disrupted, which results in the reduced capacity of the sector to respond to the primary health consequences of flooding and delivering care to COVID-19 patients (21). Hence, disaster management approaches need to account for the effect of social contact network structures, policy interventions, and compare the risk of flooding with ones of COVID-19 to prepare evacuation plans. Ishiwatari *et al.* (20) recommended that the government consider lifting requirements to encourage people to take shelters against the large scale flooding disasters, and citizens to shelter in place in case of inland flooding. We argue that given the intra-nation heterogeneities in underlying socioeconomic factors and healthcare responsiveness (3,22), risk management frameworks need to quantitatively examine the primary consequences of flooding and secondary effects of COVID-19 transmission at a local scale.

To address these conflicting objectives associated with the compound risk arising from the flood hazard and pandemic COVID-19 in designing the flood evacuation strategy, here we develop a multi-objective optimization framework. The optimization model's objectives are to reduce the number of new infections in the shelters after the shelter stay period and increase the number of flood evacuees from the villages under high flood hazards. There is also a competition among the villages to have the highest possible evacuations, where the total number of evacuees is highly constrained due to the pandemic (Figure 1(a-b)). We address these multiple objectives using the max-min approach of multi-objective optimization, which has been widely used in areas such as water resources management (23–25), waste load allocations for water quality management in a stream (26–29). The model is applied to a flood-prone district on the east coast of India, the Jagatsinghpur district in Odisha.

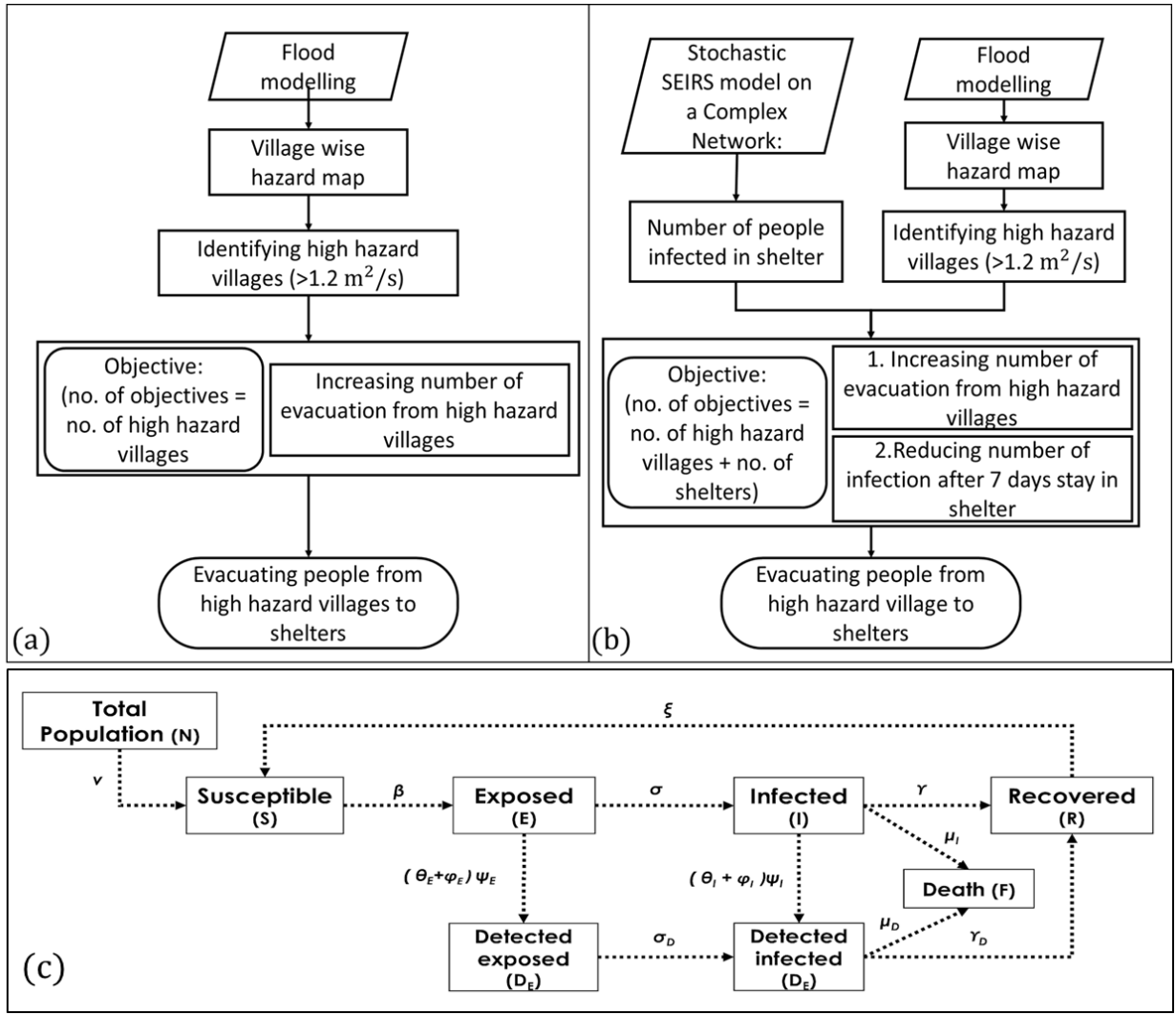

*Figure 1*: Flow chart of evacuation strategy planning: (a) Without considering the COVID-19 pandemic scenario; (b) Considering COVID-19 pandemic scenario;(c) Flowchart of SEIRS Plus model used in this study. $\nu$, $\beta$, $\sigma$, $\gamma$, $\mu$ and $\xi$ represent the rate of transmission from total population to susceptible, susceptible to exposed, exposed to infected, infected to recovered, infected to fatality state, and recovered to susceptible respectively. Parameters $\theta_E$ and $\theta_I$ are testing rates, whereas $\psi_E$ and $\psi_I$ are positivity rate for exposed and infected individuals, respectively.

## 2. Case-study and Data

Jagatsinghpur is a coastal (east coast) district in the state Odisha, India (Figure 2(a)). It comprises eight blocks, two municipalities, eight tehsils (sub-district level), 194 gram panchayats (village administrative divisions), and 1294 villages. The district covers a total area of 1759 km² and has a population of about 1.14 million, according to the Census of India (30). The four major rivers of Odisha, Mahanadi, Devi, Kathajodi, and Biluakhai pass through this district. Due to the location and geographical conditions, Jagatsinghpur is prone to riverine and coastal flooding. As a part of preparedness measures, Odisha State Disaster Management Authority (OSDMA) has built multipurpose cyclone and flood shelters (MCS and MFS respectively) at strategic locations for the vulnerable communities (Figure 2(b)). These shelters are mostly situated at locations that are not more than 2.25 km from any part of any of all villages (9). Apart from these shelters, various school buildings are also used as shelters during flood and cyclone events.

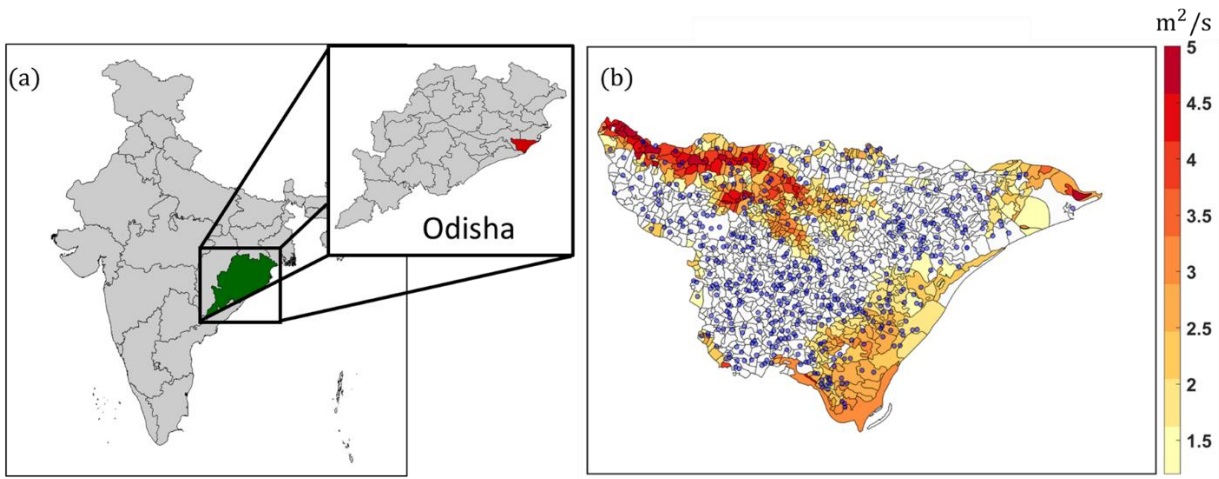

*Figure 2*: *(a) Location of Jagatsinghapur district in the state of Odisha in India; (b) Village wise hazard values in Jagatsinghpur (considering a 100 year return period and 24 hours duration flood event) and shelter locations (499 shelter locations specified by blue dots)*

The first step in designing any evacuation strategy is to identify the villages with high flood hazard. The hazard values associated with 100 years return period were estimated for the Jagatsinghapur district as reported by Mohanty et al. (31). The authors have considered regionalized design rainfall, design discharge, and design storm-tide as primary inputs to a comprehensive 1D-2D coupled MIKE FLOOD model (32) to derive flood hazard values at village level. In the present study, hazard values generated for a flood quantiles corresponding to 100-years return period and 24 hours duration are considered to classify the villages into the category of "high hazard" (31,33). Villages with hazard value above 1.2 m$^2$/s are considered as high hazard villages, where evacuation are needed to be done. We find that there are 484 villages experiencing high flood hazards, out of which, we have selected 397 villages based on the availability of most recent granular population data from the Census of India (30). The villages are marked in figure 2(b) with their respective hazard values. OSDMA, with the help of state government, central government, and World Bank, has built 21 multipurpose flood shelters (MFS) and 21 multipurpose cyclone shelters (MCS) in the district, which are situated near vulnerable areas (9). Along with these, 542 schools are also used as shelters during floods or cyclones. Out of these shelters, we consider 499 shelters based on available data from the Government data sources (http://gisodisha.nic.in/District/jagatsinghpur/; last access: 24 May 2020). The distance between each of the high hazard villages and the corresponding shelters play a major role in the evacuation operation. As it is not recommended to evacuate people to distant shelters, considering the constraints associated with transport during extremes and the evacuees' comfort levels, here we consider the five nearest shelters for each of the high flood hazard villages. According to the district emergency office, the maximum capacity of each shelter is 2000.

## 3. Model Development

### 3.1. Optimization model

The optimization model that is needed to be solved for designing evacuation strategies has the objective functions to maximize the number of evacuees from individual (likely) flood-affected villages. Hence, the number of objectives for such a model will be the same as the number of villages with high hazard values (Figure 1(a)). Under the pandemic, the optimization model for flood evacuation will further involve another set of objectives to reduce the number of likely infections after the stay period at each of the shelters. Hence, the number of objective functions for the present case under the pandemic scenario is the sum of the number of villages and the shelters. In village $i$, let us assume that the population is $pop_i$, which comprises people living in Kutcha houses (The walls and/or roof of which are made of material other than those mentioned above, such as un- burnt bricks, bamboos, mud, grass, reeds, thatch, loosely packed stones, etc. are treated as kutcha house) and Pucca houses (A pucca house is one, which has walls and roof made of the following material. Wall material: Burnt bricks, stones (packed with lime or cement), cement concrete, timber, ekra etc.). Let us assume, the fraction of people living in Kutcha houses in village $i$ is $zx_i$, and the fraction of people living in Pucca houses is $zy_i$. As more risk is associated with the people in Kutcha houses, we need to incorporate this information into the optimization model. Let us assume that the fraction of risk associated with kutcha house people in $rf_i$. The ratios of evacuees from Kutcha and Pucca houses to total population in the $i^{th}$ village are $x_i$ and $y_i$. The reduction of flood risk $R_i$ after an evacuation may be defined as $R_i = rf_i \times \frac{x_i}{zx_i} + (1 - rf_i) \times \frac{y_i}{zy_i}$. The optimization model maximizes $R_i$ for all the villages during the evacuation process. The other set of objective functions during the pandemic is to minimize the number of infections $(I_j)$ in shelter $j$ after the stay. $I_j$ is a function of the number of evacuees in the shelter $(E_j)$ and the total number of initial infections $(I_{j,0})$. $e_{ij}$ denotes the number of evacuees from $i^{th}$ village to $j^{th}$ shelter.

Finally, the resulting optimization model is expressed as:

$$Maximize\ R_i \quad \forall\ i \tag{1}$$

$$\text{Minimize } I_j \quad \forall j \quad (2)$$

$$0 \leq (x_i + y_i) \leq 1 \quad \forall i \quad (3)$$

$$R_i = rf_i \times \frac{x_i}{zx_i} + (1 - rf_i) \times \frac{y_i}{zy_i} \quad \forall i \quad (4)$$

$$(x_i + y_i) \times pop_i = \sum_j e_{ij} \quad \forall i, j \quad (5)$$

$$E_j = \sum_i e_{ij} \quad \forall i, j \quad (6)$$

$$I_j = f(E_j, I_{j,0}) \quad \forall j \quad (7)$$

$$0 \leq x_i \leq zx_i \quad \forall i \quad (8)$$

$$0 \leq y_i \leq zy_i \ (if \ x_i < zx_i, \ y_i = 0) \quad \forall i \quad (9)$$

$$0 \leq E_j \leq E_{j,max} \quad \forall j \quad (10)$$

$$e_{ij} = 0 \quad \forall j \notin S_i \quad (11)$$

Where, $E_{j,max}$ is the capacity of $j^{th}$ shelter, which is considered here to be 2000, as per the information provided by Government agency. $E_j$ is the number of evacuees staying is shelter $j$. $S_i$ is the set of shelters that belong to the five closest shelters from village $i$. The function $f$ is Eq. (7) is an epidemiological model based on the extended Susceptible - Exposed - Infectious - Recovered – Susceptible (SEIRS) model. Eq. (9) takes care of the fact that the people in pucca houses will be evacuated only after the complete evacuation of the population living in the kutcha houses.

### 3.2. SEIRS Epidemiological Model

To study the effect of social contact network structures on the propagation of the spread of COVID-19 (SARS-CoV-2) in the community as a consequence of community gathering in the shelter house, we use extended SEIRS model. In the standard SEIRS model, the entire population is divided into Susceptible (S), Exposed (E), Infectious (I), and Recovered (R) individuals. In the extended SEIRS model, we further divide the population in Detected Exposed (DE) and Detected Infected (DI) by using social tracing and testing parameters. The initial seed is then provided in terms of population in each category. Recent developments in the field of epidemiological modeling further compartmentalize the contagious individuals according to the degree of severity of symptoms. However, given the limited availability of datasets to calibrate the associated parameters, we use a 7-compartment model in this study (Figure 1(c)).

A Susceptible member becomes exposed or infected upon contacting the infected individual during a transmission event. Newly exposed individuals experience a latent period during which they are not contagious (referred to as the Incubation period). Exposed individuals than progress to the infected stage where they can either get tested if they are exhibiting symptoms or they have been selected for testing based on the contact tracing network at the prevalent rate of contact tracing and testing in the society. The infected individual can then progress either to Recovery (R) or succumb to the infection (F).

Since we are interested in decreasing the flood risk in the COVID-19 scenario, we use the deterministic mean-field model implementation of the SEIRS Extended model. Specifically, we assume that despite the underlying interaction social interaction structure that is ubiquitous to any society, the interactions within the shelter homes will primarily be random due to the violation of social distancing norms. Hence, all individuals mix uniformly and have the same rates and parameters in the current implementation of the epidemiological model. We use the SEIRSPLUS package implemented in Python to obtain the number of infected individuals in each shelter filled with the full capacity of 2000 using different values for the initial number of infections. We note that if the underlying social network structure and information on testing and isolation testing protocols are available, stochastic network models are recommended to account for stochasticity, heterogeneity, and deviations from uniform mixing assumptions (34).

### 3.3. Max-Min Approach

The multi-objective optimization model presented in Eq. (1-11) is solved here with the Max-min approach. The first objective function can have a value between 0 to 1 as per Eq. (4). The second objective function is also standardized by dividing $I_j$ by $I_{j,max}$, which is the maximum possible value of $I_j$, given by, $f(E_{j,max}, I_{j,0})$. The max-min approach maximizes the minimum of all the objectives (when objectives are to be minimized, it is considered as the maximization of the negative of the objective function), which will force all the individual objectives to maximize. Following the max-min approach, the optimization model may be formulated as:

$$\text{Maximize } \lambda \tag{12}$$

$$R_i \geq \lambda \quad \forall i \tag{13}$$

$$1 - \frac{I_j}{I_{j,max}} \geq \lambda \quad \forall j \tag{14}$$

$$0 \leq (x_i + y_i) \leq 1 \quad \forall i \tag{15}$$

$$R_i = rf_i \times \frac{x_i}{zx_i} + (1 - rf_i) \times \frac{y_i}{zy_i} \quad \forall i \tag{16}$$

$$(x_i + y_i) \times pop_i = \sum_j e_{ij} \quad \forall i, j \tag{17}$$

$$E_j = \sum_i e_{ij} \quad \forall i, j \tag{18}$$

$$I_j = f(E_j, I_{j,0}) \quad \forall j \tag{19}$$

$$I_{j,max} = f(E_{j,max}, I_{j,0}) \qquad \forall j \qquad (20)$$

$$0 \leq x_i \leq zx_i \qquad \forall i \qquad (21)$$

$$0 \leq y_i \leq zy_i \ (if \ x_i < zx_i, \ y_i = 0) \qquad \forall i \qquad (22)$$

$$0 \leq E_j \leq E_{max} \qquad \forall j \qquad (23)$$

$$e_{ij} = 0 \qquad \forall j \notin S_i \qquad (24)$$

$$0 \geq \lambda \geq 1 \qquad (25)$$

The model mentioned above is a non-linear optimization model. We use a search algorithm, known as Probabilistic Global Search Laussane (PGSL) to obtain the feasible optimal solution (35).

### 3.4. PGSL: Search Algorithms for Optimization Model

PGSL, a global search algorithm, was developed by Raphael and Smith (35) based on the assumption that better results can be obtained by focusing more on the neighborhood of good solutions. In every iteration, the algorithm increases the probability of obtaining a solution from the region of good solutions of the previous iteration. Thus, the search space is narrowed down until it converges to the optimum solution. PGSL is different from other methods as it uses four nested cycles, which helps improve the search, and thus more focus could be given to areas around good solutions (29). The four cycles of PGSL are:

Sampling cycle: Samples are generated randomly from the current PDF of each variable. Each point is evacuated based on the objective functions, and the best point is selected.

Probability updating cycle: Probability of neighborhood of good results increased and bad decreases, and the PDFs of each variable are updated accordingly after each cycle.

Focusing cycle: Search is focused on an interval containing better solutions after a number of probability updating cycles. This is done by dividing the interval containing the best solution for each variable.

Subdomain cycle: The search space keeps narrowing by selecting only a subdomain of the region of good points.

## 4. Results and discussion

We apply the developed optimization model (Eq. 12-25) to the case study of Jagatsinghapur District. Due to data non-availability, we have assumed hypothetical values of some of the variables for demonstration purposes. We considered *zx* to be 0.6, *zy* to be 0.4 and *rf* to be 0.8 for all the villages (i). The maximum shelter capacity is considered to be 2000. The stay period in the shelter is considered to be seven days. The values are considered after discussions with planners and management authorities working at different levels of decision making. We applied our optimization model first by considering a uniform initial infection value across the district. The infection value is considered as 1% for demonstration purpose. We first simulated the increase in the number of infections in a shelter assuming different initial infection values (0.1%, 0.25%, 0.5%, 0.75%, and 1%) with shelters at full capacity for seven days. We find, the number of infections in a shelter to be in the range of 7 to 60 at the end of the stay, depending on the initial infection (Supplementary figure S1). Violation of social distancing norms within shelter houses operating at their designated capacity could expose a large number of individuals to the highly contagious diseases, including COVID-19. Once exposed and infected individuals move back to their respective villages, it may result in the widespread outbreak at local scales.

Such a scenario may also become unmanageable, as the medical, as well as other facilities, will be limited after flooding events. Hence, a proper evacuation strategy planning is needed to decrease flood losses and the spread of COVID-19.

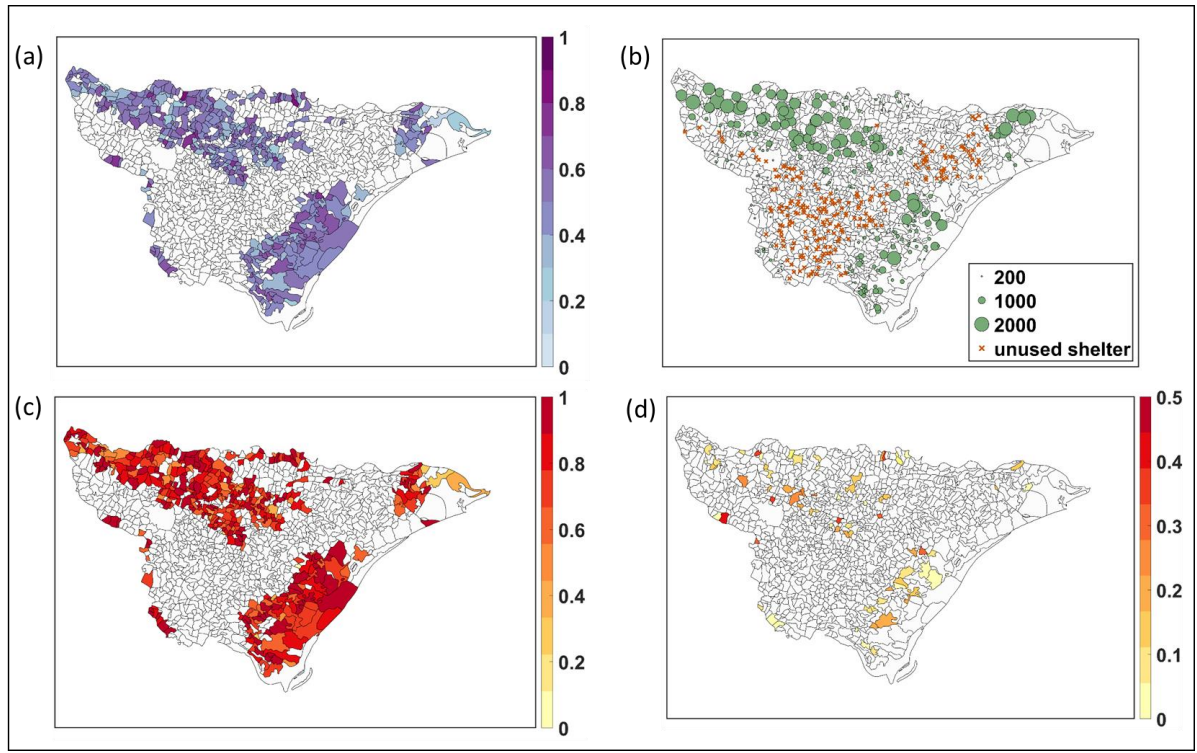

**Figure 3**: *Results from the optimization model without considering the COVID-19 pandemic situation: (a) Fraction of total population evacuated; (b) Population of shelter houses (213 unused shelters specified by the red cross); (c) Fraction of kutcha house people evacuated; and (d) Fraction of pucca house people evacuated*

To check the applicability of our optimization model, we first used the optimization model for non-pandemic scenarios, which includes the equations from Eq (12) to Eq (25), excluding Eq (14), (19), and (20). The results obtained from the model are presented in figure 3. We find that in most of the villages with high flood hazards, more than 50% of the population is evacuated (Figure 3(a)). We find a good number of shelters (213) remain unused in the Central area (Figure 3(b)), as they are far away from the hazardous villages, and transporting people to those shelters is difficult. These shelters may not be useful during the flood, but during cyclones, they are extensively used. In most of the villages considered, more than 75% of the populations in kutcha houses are getting evacuated (Figure 3(c)). For 69 villages, this fraction reaches 100% with evacuations for a few in pucca houses (Figure 3(d)). Such realistic results prove that the model works efficiently under non-pandemic situations.

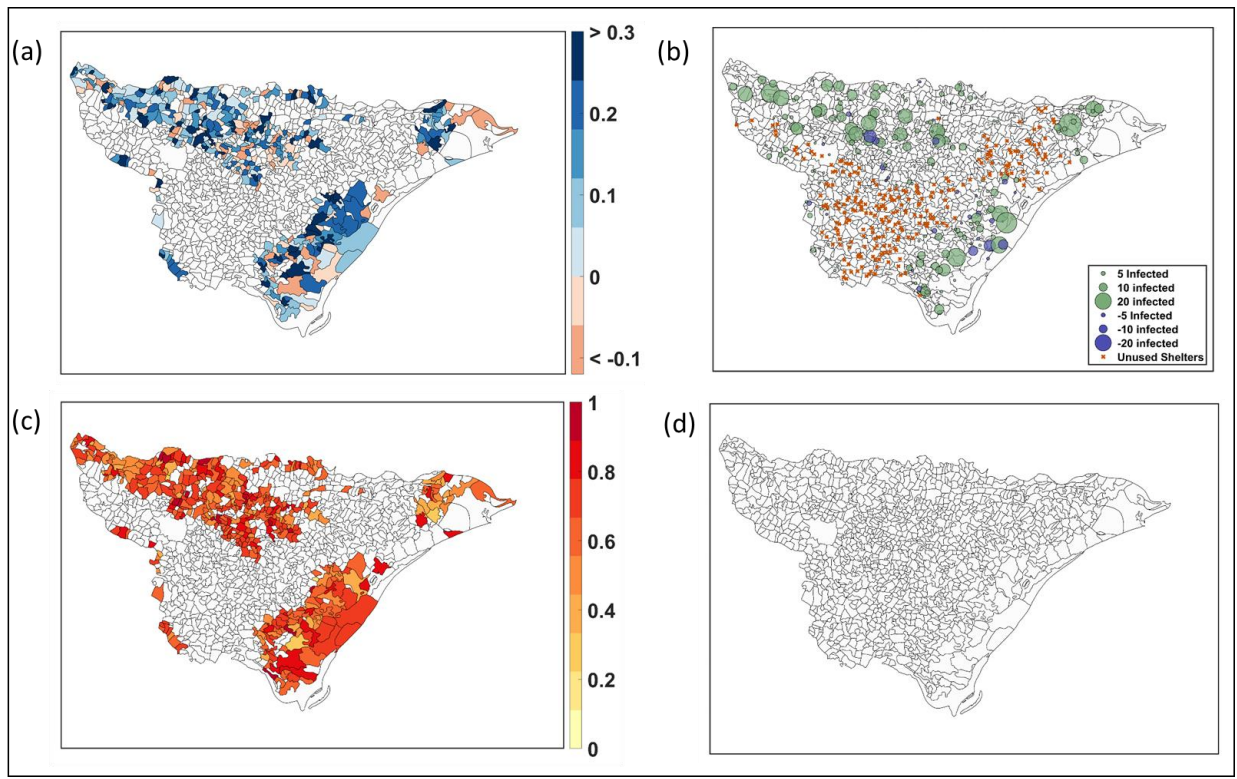

*Figure 4: (a) Difference in the fraction of total population evacuated between without and with COVID-19 scenario; (b) Difference in the number of infected people between the cases, without and with considering COVID-19 risk in the optimization, for each shelter at the end of 7 days; (c) Fraction of kutcha house people evacuated with the COVID-19 condition; and (d) Fraction of pucca house people evacuated with the COVID-19 condition. Note, these values remain zero at all the villages*

To apply the model during the pandemic, we first considered the initial infection to be 1% of the population uniform across the district. Consideration of pandemic reduces the number of evacuees, which is evident in figure 4(a). There are a few villages where the numbers of evacuees have increased marginally, resulting from multiple possible solutions for the multi-objective optimization model. The existence of multiple solutions is quite common and is observed in multiple other applications (29,36). However, despite the existence of multiple solutions while handling around 900 objective functions (sum of the number of shelters and number of villages) and 4840 decision variables (484 villages, with the number of evacuees from kutcha and pucca houses to nearest five shelters), the model shows an overall decrease in the number of infections after the shelter-stay period (Figure 4(b)), when compared to the case if evacuation planning, as presented in figure 3, would have been followed. In most of the shelters, the number of infections reduces by good numbers after considering the pandemic related objectives; though there are a few with a slight increase in infections (less than ten). They essentially result from the existence of multiple solutions. In some of the shelters, the number of infections has been reduced by more than 20, showing the effectiveness of our model. We observe that the proposed model is that it still evacuates more than 80% of the population from Kutcha houses, even during the pandemic, and does not evacuate people from the *pucca* houses as these are relatively safer than the Kacha houses. Such as assignment of priority makes the model effective to the Indian coastal regions.

One of the limitations of the proposed model is that we have not assigned higher weights to the objective functions related to evacuations of villages under very high flood hazard. To overcome this, we have multiplied the LHS of Eq (13), $R_i$ by a factor (= *1.2 / hazard value*), and this will reduce the value of LHS in the same equation. Lowering the value of an objective will increase its importance and in the max-min approach, as we are maximizing the minimum of all the objectives.

The results obtained for the non-COVID19 scenario, with the consideration of weights (Supplementary figures S2 and S3 (a)) show that the number of evacuees is quite high in the villages with very high and extremely high flood hazards (Supplementary Table ST1). We could not achieve such results inversely associated with the hazard values using the optimization without the assignment of weights. However, under the COVID19 scenario, the objectives associated with minimizing the infections do not allow the number of evacuees to reach high values even in villages with high flood hazards (Supplementary figures S3 (b) and (c)), and it is almost the same to the number of evacuations for the no weight case. Consideration of the very high number of objectives in such cases also dilutes the impacts of weights.

To understand the model applicability for more realistic cases, we have also considered different initial infections in different villages with high flood hazard. We have classified the villages into two categories based on their geographic locations; coastal and interior (Figure 5 (a)). We have considered the optimization results for four cases, I: no COVID scenario; II: The initial infection rate at coastal villages is 0.1%, and at interior villages 1%; and III: Initial infection rate at coastal villages is 1% and at interior villages 0.1%. Figure 5 (b) shows that the model reduces the number of evacuations from the high infected villages significantly. The higher difference in case II (or III) from the case I show lower evacuations. However, even with a lower evacuation, there are quite a high number of infection at the end of evacuation periods in the shelters situated in high infection regions. This also poses another issue of mixing people from two villages of very different infection rates in a shelter. Presently the model does not consider this criterion with an assumption that villages with similar infection rates are situated at similar locations. The handling of villages with differential infection rates may be considered as the potential area of future research.

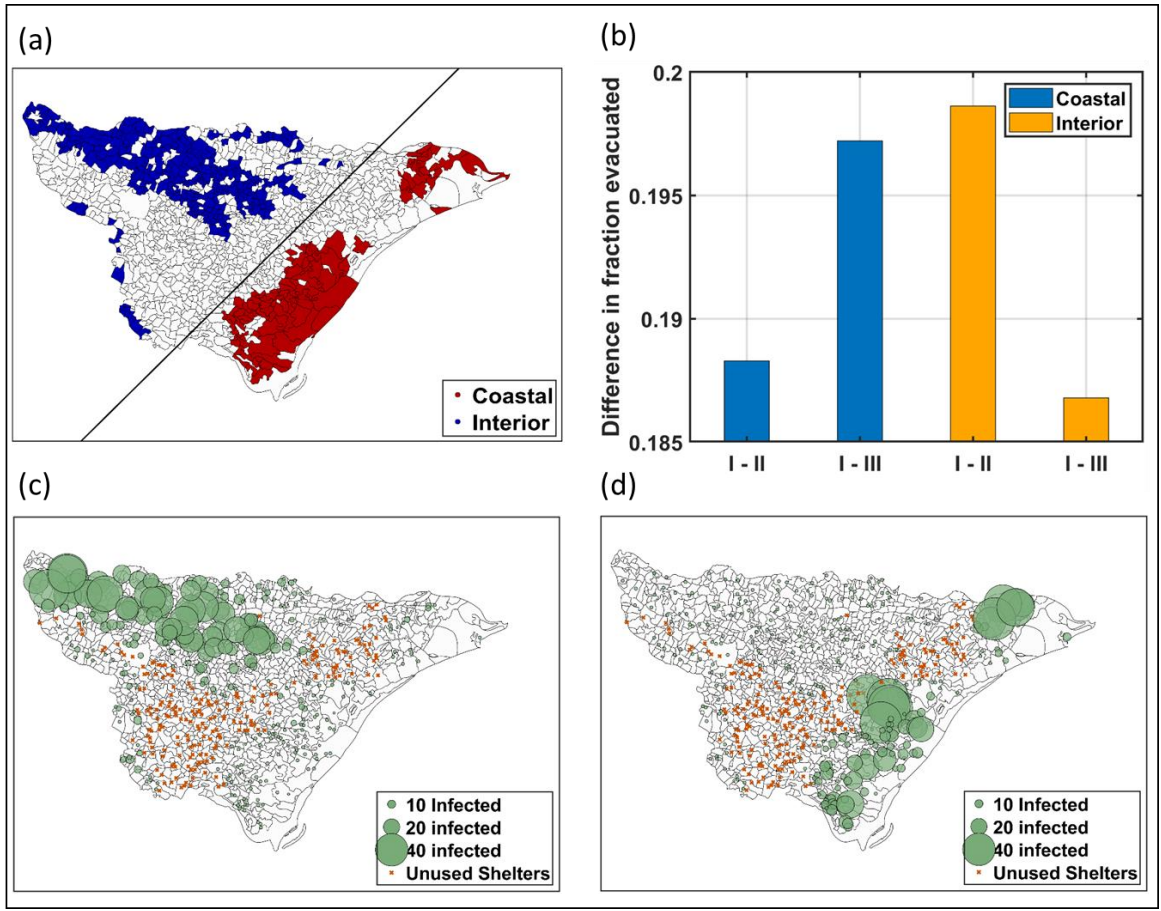

*Figure 5*: *(a) Classification of villages into coastal and interior; (b) Differences in fractions of evacuation between case I & case II and case I & case III in coastal and interior villages (c) People infected in each shelter in case II; and (d) People infected in each shelter in case III.*

*Case I: No COVID scenario, Case II: Initial infection rate at coastal villages is 0.1% and at interior villages 1%, Case III: Initial infection rate at coastal villages is 1% and at interior villages 0.1%.*

## 5. Summary

Here we address an important compound event problem related to flood evacuation to minimize the loss due to flood hazards, while also demonstrating the potential issues associated with evacuation processes, when floods co-occur with a pandemic. The handling of conflicting objectives set by multiple players makes the problems complex. The players of this problem are categorized into two groups: an increase in number of evacuations in each village and a decrease in number of infections in each shelter. This makes the number of objectives to be equal to the sum of the number of villages and the number of shelters. These multiple objectives are handled with max-min approaches. The proposed model follows the simulation-optimization approach, where the simulations of the spread of infections are taken care by the epidemiological model, while the optimization model identifies the evacuation strategy. Results show that the model effectively handles a number of objectives by reducing the number of evacuations from the villages with higher infections. Though an increase in infections is inevitable post-evacuation, it is possible to restrict the maximum infections per shelter to a count of 40 in most cases. This may be considered as a condition under control, given that the area per head allotted in these shelters is small, which makes it impossible to follow the norm of social distancing. The proposed algorithm is an example of flood evacuation strategy, and the same can be extended for cyclone evacuation as well.  A major limitation of the model, like any other multi-objective optimization model, is the existence of multiple solutions. However, often the existence of multiple solutions is preferred by a policymaker, as they provide multiple options for decision-making in real-life critical scenarios.

**Supplementary Information for**

**Flood Evacuation During Pandemic: A multi-objective Framework to Handle Compound Hazard**

Shrabani S. Tripathy[1], Bhatia Udit[2], Mohit Mohanty[3], Subhankar Karmakar[1, 3, 4], Subimal Ghosh[1, 4, 5, *]

[1]Inter-Disciplinary Program in Climate Studies, Indian Institute of Technology Bombay, Mumbai – 400076, India

[2]Dept. of Civil Engineering, Indian Institute of Technology, Gandhinagar, Palaj – 382355, India

[3]Environemntal Science and Engineering Department, Indian Institute of Technology Bombay, Mumbai – 400076, India

[4]Centre for Urban Science and Engineering, Indian Institute of Technology Bombay, Mumbai – 400076, India

[5]Department of Civil Engineering, Indian Institute of Technology Bombay, Mumbai – 400076, India

[*]Correspondence to Subimal Ghosh, Email: subimal@civil.iitb.ac.in

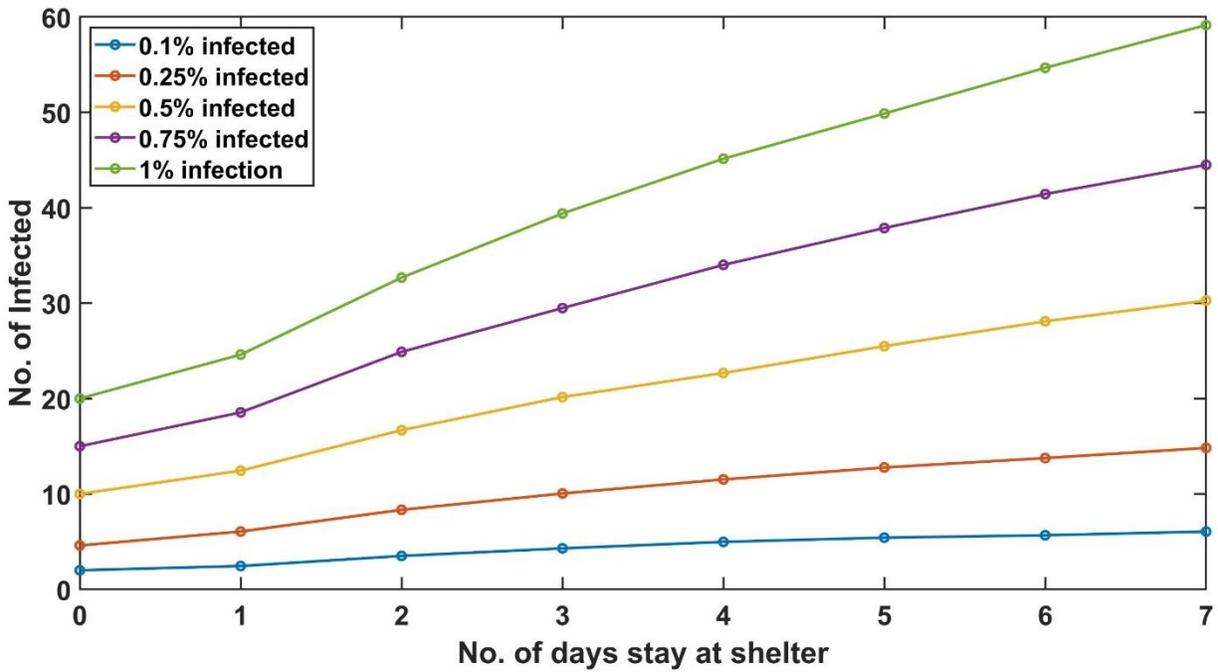

***Supplementary figure S1***: *Day wise number of infected in each shelter (when shelter is filled to full capacity, 2000) depending upon initial number of infection.*

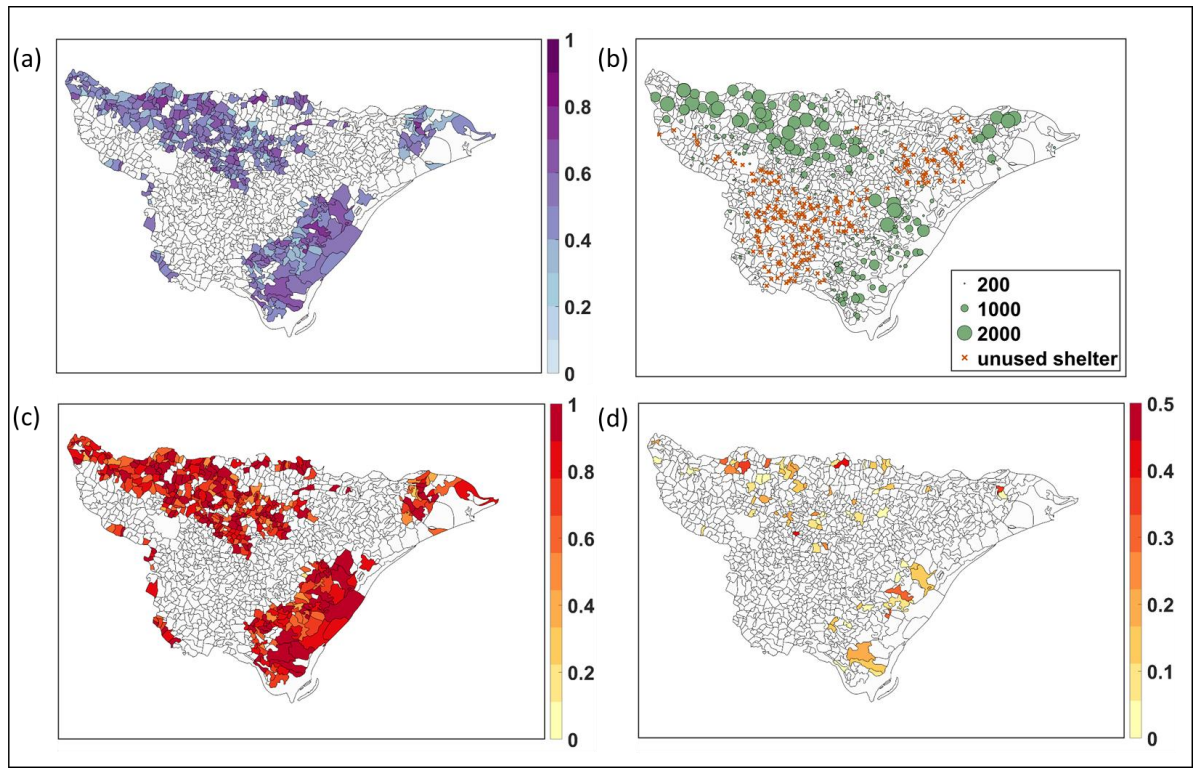

***Supplementary figure S2***: Results from the optimization model without considering the COVID-19 pandemic situation (after considering weight related to flood hazard magnitude): *(a) Fraction of total population evacuated; (b) Population of shelter houses(213 unused shelter specified by red cross); (c) Fraction of kutcha house people evacuated; and (d) Fraction of pucca house people evacuated*

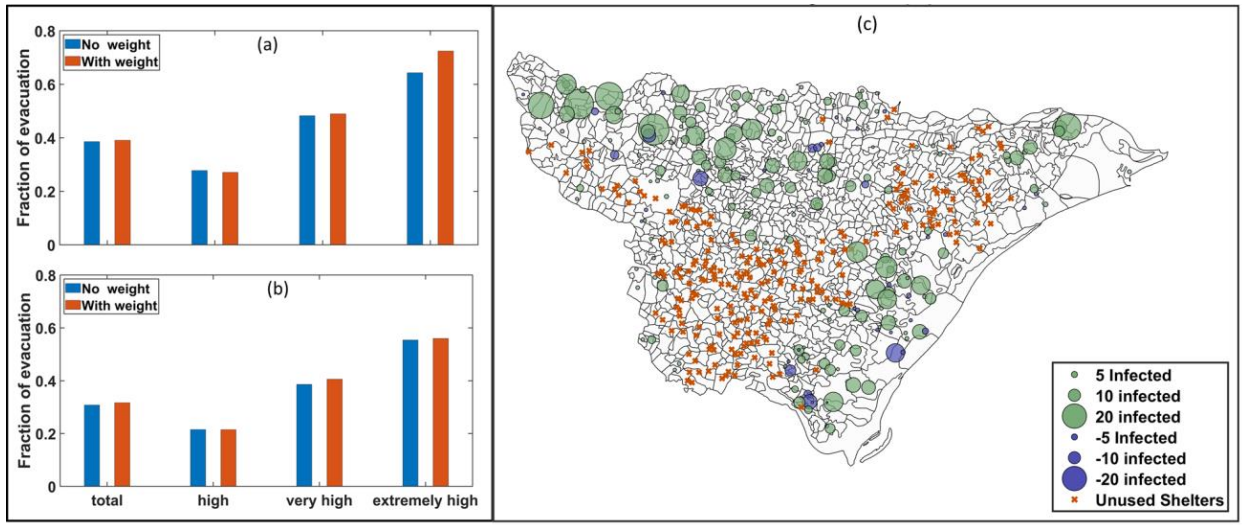

***Supplementary figure S3***: *(a)Fraction of people evacuated under no COVID scenario (with and without considering weights related to flood hazard magnitude); (b)Fraction of people evacuated from different category of villages under COVID scenario (with and without considering weights related to flood hazard magnitude); (c)Difference in the number of infected people with and without considering COVID-19 scenario in each shelter at the end of 7 days (considering weights related to flood hazard magnitude)*

**Supplementary table ST1**: Hazard categories of villages

| Categories | Flood Hazard (m$^2$/s) | No. of villages |
|---|---|---|
| **High** | 1.2 - 1.8 | 104 |
| **Very high** | 1.8 - 3.6 | 242 |
| **Extremely high** | > 3.6 | 51 |